\documentclass{article}
\usepackage{epsfig}
\usepackage{graphicx}

\DeclareGraphicsExtensions{.pdf,.png,.jpg}

\begin{document}

\title{Dipolar Radicals in Crossed Electric and Magnetic Fields}

\author{John L. Bohn \\
\small JILA, Department of Physics and NIST,
University of Colorado, Boulder, CO 80309 \\
Goulven Qu\'em\'emer \\
\small Laboratoire Aim\'e Cotton, Universit\'e Paris-Sud, CNRS,
B\^atiment 505, 91405 Orsay, France.}

\maketitle

\abstract{ Paramagnetic, dipolar Hund's case-a radicals are considered in the presence of arbitrary, non-collinear combinations of electric and magnetic fields. The field-dependent part of the Hamiltonian is found to be
exactly diagonalizable, and described by quantum numbers given by the projection of the molecule's total angular
momentum along a space-fixed axis that is determined by both the fields and the electric and magnetic dipole moments of the molecule.  In cases of strong fields, this procedure identifies a set of quantum numbers
for the molecule in crossed fields.  We dub this set a ``Hund's case-X'' basis.}

\section{Introduction}

The concept of Hund's angular momentum coupling cases is useful for naming and organizing the energy levels of molecules.  Quite generally, the molecular Hamiltonian can be written as a sum of several pieces, $H = H_1 + H_2 + H_3 + \dots$, which may fail to commute among themselves.  Hence the eigenstates of $H$ cannot be labeled by the quantum numbers appropriate to each $H_i$ simultaneously.  However, often one of the terms, let us say $H_1$, dominates over the others.  It is then worthwhile to express eigenstates of $H$ in terms of the eigenstates and quantum numbers of $H_1$, whereby contributions off-diagonal in these states, arising from $H_2, H_3, \dots$ are perturbative. In this way, while the quantum numbers of $H_1$ are not strictly ``good'' quantum numbers, they are ``good enough'':  they serve to classify the states, identify characteristic energy level spacings, and provide approximate line strengths for transitions \cite{BC}.

For a molecule immersed in either an external electric or magnetic field, one such quantum number is the projection $m$ of the molecule's total angular momentum onto the field axis.  This is in fact a rigorously good quantum number, and one that describes the joint system of molecule-plus-field.  However, a molecule that is both dipolar and paramagnetic, such as OH, can respond to both electric and magnetic fields.  If these fields are not collinear, then rotational invariance of the Hamiltonian is broken, and neither the $m$ quantum number referred to the electric field axis, nor the one referred to the magnetic field axis, remains good.

Our main point here is the following.  For a molecule, like OH, which is represented by Hund's case-a in the absence of fields, both the electric ${\vec d} = d{\hat n}$ and magnetic ${\vec \mu} = \mu {\hat n}$  moments can be regarded, to a good approximation, as collinear with the molecular axis ${\hat n}$.  In combined electric ${\vec {\cal E}}$ and magnetic ${\vec {\cal B}}$ fields, the field Hamiltonian consists of Stark and Zeeman terms,
\begin{eqnarray}
\label{Field_Hamiltonian}
H_{\rm field} = H_{\rm S} + H_{\rm Z} &=& -{\vec d} \cdot {\vec {\cal E}} - {\vec \mu} \cdot {\vec {\cal B}}
\\
&=& -{\hat n} \cdot \left( d {\vec {\cal E}} + \mu {\vec {\cal B}} \right). \nonumber
\end{eqnarray}
Geometrically, this Hamiltonian describes a generalized ``moment'' ${\hat n}$ interacting with a ``combined field'' that is a weighted linear combination of the electric and magnetic fields.  The combined field determines an axis of symmetry with respect to which meaningful $m$ quantum numbers can again be assigned.  These are the
``good enough'' quantum numbers in this situation, and define therefore a kind of Hund's case, useful even when
other effects such as $\Lambda$-doubling are considered.

In the following we elaborate on this idea, showing various examples for the OH molecule
and the di-halogen ICl.  Understanding the behavior of OH in crossed electric and magnetic fields has suddenly increased in importance, given recent experiments in which trapped gases of this radical, at mK temperatures, experience widely varying relative magnitudes and orientations of the fields \cite{Stuhl12_PRL,Stuhl12_Nature}.

\section{Formulation}

We begin with a Hund's case-a molecule in a field, described in a given electronic state by the effective Hamiltonian
\begin{eqnarray}
H = H_{\rm SO} + H_{\rm rot} + H_{\rm S} + H_{\rm Z} + H_{\Lambda} + \dots,
\end{eqnarray}
which represent, in order, the spin-orbit, rotatational, fields, and $\Lambda$-doubling contributions.  We will assume all other effects are perturbative and can be included as necessary.  Hund's case-a brings with it not only a set of quantum numbers, but also a hierarchy of quantum numbers.  The primary ones are those pertinent to $H_{\rm SO}$, namely, the signed projections of electronic orbital ($\lambda$) and spin ($\sigma$) angular momenta, and their sum $\omega = \lambda + \sigma$. We will assume throughout that these quantum numbers are well-defined in the electronic state of interest, e.g., the $^2 \Pi_{3/2}$ ground state of OH or the $A^1 \Pi_1$ state of ICl.

Given the value of $\omega$, a secondary quantum number $j$ describes the rotational eigenstates generated by $H_{\rm rot}$.  The value of $j$ cannot be set independently of $\omega$, but is rather {\it contingent} on $\omega$, since it must satisfy $j \ge |\omega|$.  Finally, for a given value of $j$, the value of $m$ is contingent on both the value of $j$, via the usual restriction $-j \le m \le j$, {\it and} on the space-fixed axis used to quantize the angular momentum, which gives it  a concrete meaning.  As alluded to above, this choice of axis is usefully specified by the direction of a single field.  Thus the Hund's case-a basis set is indexed by a particular collection of meaningful quantum numbers:
\begin{eqnarray}
|\lambda \sigma \rangle |\omega j m\rangle,
\end{eqnarray}
where the first ket describes the electronic degrees of freedom in the body frame, and the second ket describes the distribution of the molecular orientation in this frame, via
\begin{eqnarray}
|\omega j m  \rangle = \sqrt{ \frac{ 2j+1 }{ 8\pi^2 }}
D^{j*}_{m \omega}(\alpha \beta \gamma),
\end{eqnarray}
where $(\alpha \beta \gamma)$ are the Euler angles relating the molecular axis ${\hat n}$ to the laboratory-fixed quantization axis.

As an aside, we note that this basis can be transformed so as to diagonalize the $\Lambda$-doublet Hamiltonian $H_{\Lambda}$, by constructing parity eigenstates that are linear combinations of the states $|+\omega \rangle$ and $|-\omega \rangle$.  However, as we are mostly concerned here with states in large electric fields, it is more appropriate to use the states with signed values of $\omega$.  To this end, we will denote the magnitude of $\omega$ as ${\bar \omega} = |\omega|$ where necessary.

\subsection{Electric fields}

The molecule is assumed to be polar, with electric dipole moment ${\vec d}
= d {\hat n}$, where ${\hat n}$ denotes the molecular axis.  In an electric field
${\vec {\cal E}}$ the molecule experiences a Stark energy
\begin{eqnarray}
H_{\rm S} &=& -{\vec d} \cdot {\vec {\cal E}} \nonumber \\
&=& -d{\cal E} \cos (\beta),
\end{eqnarray}
where ${\cal E} \cos (\beta )$ is the projection of the field on the molecule's axis,
and $\beta$ is the angle between field and dipole.
This operator has no explicit dependence on electron coordinates, and so the electronic matrix element is unity.  Moreover, let us consider a field sufficiently weak that the angular momentum $j$ is nearly conserved.  The matrix elements of the Stark Hamiltonian are then
\begin{eqnarray}
\label{Stark}
\langle \lambda \sigma | \langle \omega j m | H_{\rm S}
|\omega  j m \rangle |\lambda \sigma \rangle &=& -d {\cal E}
\langle \lambda \sigma | \lambda \sigma \rangle
\langle \omega  j m | \cos (\beta) |\omega  j m \rangle \nonumber \\
&=& -d {\cal E}
\langle \omega j m | \cos (\beta) |\omega j m  \rangle
\end{eqnarray}
The remaining matrix element has a standard form \cite{BS}:
\begin{eqnarray}
\langle \omega^{\prime} j m^{\prime}  | \cos (\beta) |\omega  j m \rangle =
(-1)^{m^{\prime} - \omega^{\prime}} (2j+1)
\left( \begin{array}{ccc} j & 1 & j \\
                          -\omega^{\prime} & 0 & \omega \end{array} \right)
\left( \begin{array}{ccc} j & 1 & j \\
                          -m^{\prime} & 0 & m \end{array} \right).
\end{eqnarray}
This expression is of course diagonal in $m$, for the quantization axis ${\hat {\cal E}}$.
It is also diagonal in $\omega$, reminding us that the signed value of $\omega$ is a good quantum number in the presence of the field.  Indeed, re-writing this matrix element in terms of the Wigner-Eckart theorem,
\begin{eqnarray}
\label{WE}
\langle \omega j m | T^1_0({\hat n}) |\omega  j m \rangle
= (-1)^{j-m} \sqrt{ 2j+1 } \langle \omega j || T^1({\hat n}) || \omega j \rangle
\left( \begin{array}{ccc} j & 1 & j \\
                          -m & 0 & m \end{array} \right),
\end{eqnarray}
identifies the reduced matrix element as
\begin{eqnarray}
\langle \omega j || T^1({\hat n}) || \omega j \rangle
&=& (-1)^{\omega - j} \sqrt{2j+1}
\left( \begin{array}{ccc} j & 1 & j \\
                          -\omega & 0 & \omega \end{array} \right),
\end{eqnarray}
where $\cos ( \beta) $ is expressed explicitly as the zero-th component of a first-rank
tensor operator $T^1_0({\hat n})$.
Substituting formulas for the 3-$j$ symbols, the reduced matrix element becomes \cite{BS}
\begin{eqnarray}
\langle \omega j || T^1({\hat n}) || \omega j \rangle
&=& \frac{ \omega }{ \sqrt{ j(j+1) } }\nonumber \\
&=& \cos( {\hat n} \cdot {\hat j}).
\end{eqnarray}
In this last line, we take the semiclassical approach, and identify this quantity
as the mean angle between the molecular axis and the total angular momentum.  The quantum number $\omega$ thus identifies the direction of the dipole moment relative to the molecule's total angular momentum.
Doing the same for the electric field factor, we can write Stark matrix elements in the form
\begin{eqnarray}
H_{\rm S} = - \left( d \frac{  \omega }{ \sqrt{j(j+1)} } \right)
\left( {\cal E}\frac{  m }{ \sqrt{j(j+1)} } \right).
\end{eqnarray}
This expression factors into a part that depends on the internal workings of
the molecule (including the dipole moment), and a part that depends on its
relation with the external field.  This Hamiltonian is diagonal in this basis,
{\it provided} that $m$ refers to quantization along the field axis ${\hat {\cal E}}$.

\subsection{Magnetic fields}

Similarly, a case-a molecule with electronic spin will experience a Zeeman shift in a magnetic field, given by
\begin{eqnarray}
H_{\rm Z} = -{\vec \mu} \cdot {\vec {\cal B}}
\end{eqnarray}
where the magnetic moment is given (in Hund's case-a) in the body frame
of the molecule as
\begin{eqnarray}
{\vec {\mu }} = -\mu_0( {\vec \lambda} + 2{\vec \sigma}),
\end{eqnarray}
where $\mu_0$ is the Bohr magneton, and ${\vec \lambda}$ and ${\vec \sigma}$ can
in principle point in any direction.  However, in a good Hund's case a)
molecule, these vectors have vanishing (or very small) contributions
in directions orthogonal to the molecular axis.  Therefore, in the $|\lambda \sigma \rangle$
electronic basis, they are replaced by their quantum numbers, and the magneitc moment
is assumed to lie parallel to the molecular axis.

As before, the Hamiltonian then depends on the projection of magnetic field on the molecular axis,
${\cal B} \cos (\beta )$ where now $\beta$ is the angle
between the molecular axis and the magnetic field.
The matrix elements of the Zeeman Hamiltonian are therefore
\begin{eqnarray}
\langle \lambda \sigma | \langle \omega j m | H_{\rm Z}
| \omega j m \rangle |\lambda \sigma \rangle
&=& \mu_0 {\cal B} \langle \lambda \sigma | (\lambda + 2 \sigma) |\lambda \sigma \rangle
\langle \omega j m | \cos(\beta) | \omega j m \rangle \nonumber \\
&=& (\lambda + 2 \sigma) \mu_0 {\cal B}
\langle \omega j m | \cos(\beta) | \omega j m \rangle .
\end{eqnarray}
This has exactly the same form as (\ref{Stark}), but with a dipole moment that depends on the values of $\lambda$ and $\sigma$, which modify the reduced matrix element.  This is the sense in which, in case-a, both dipoles
are proportional to ${\hat n}$, with the proportionality constant being
simply a reduced matrix element that expresses details (e.g., electric or magnetic)
inside the molecule.  The act of orienting ${\hat n}$ with respect to the external
field is one of pure geometry, and described by the matrix element of $\cos (\beta)$.
Proceeding as above, we can factor the energy into internal and external pieces:
\begin{eqnarray}
\label{Zeeman}
H_{\rm Z} = \left( \mu_0(\lambda + 2\sigma) \frac{  \omega }{ \sqrt{j(j+1)} } \right)
\left( {\cal B}\frac{  m }{ \sqrt{j(j+1)} } \right).
\end{eqnarray}
This Hamiltonian is diagonal in this basis, {\it provided}
that $m$ refers to quantization along the field axis ${\hat {\cal B}}$.

\subsection{Crossed fields}

Suppose now that the molecule experiences both electric and magnetic fields,
which may point in different directions.  The field part of the Hamiltonian reads
\begin{eqnarray}
\label{field_Hamiltonian}
H_{\rm S} + H_{\rm Z} = -{\vec d} \cdot {\vec {\cal E}} - {\vec \mu} \cdot {\vec {\cal B}}.
\end{eqnarray}
There is now no obvious quantization axis -- or is there?  In a particular electronic state $|\lambda \sigma \rangle$, the field Hamiltonian in the molecular orientation degree of freedom is
\begin{eqnarray}
\langle \lambda \sigma | (H_{\rm S} + H_{\rm Z}) | \lambda \sigma \rangle
= -d ({\hat n} \cdot {\vec {\cal E}})
+ (\lambda + 2 \sigma) \mu_0 ({\hat n} \cdot {\vec {\cal B}}).
\end{eqnarray}
This expression is now conveniently re-written as
\begin{eqnarray}
\langle \lambda \sigma | (H_{\rm S} + H_{\rm Z}) | \lambda \sigma \rangle
= - {\hat n} \cdot \left( d{\vec {\cal E}} - (\lambda + 2 \sigma) \mu_0{\vec {\cal B}} \right).
\end{eqnarray}
This expression has exactly the form of the dot product between an effective ``moment'' ${\hat n}$ -- identifying
the orientation of the molecule --
and an effective field that combines the electric and magnetic fields.  There are actually two such combined fields, according to the sign of the zeroth-order $g$-factor,
$g = \lambda + 2 \sigma$.  Thus there are two distinct
field Hamiltonians
\begin{eqnarray}
H_{\kappa } = -{\hat n} \cdot {\vec {\cal C}}_{\kappa},
\end{eqnarray}
where
\begin{eqnarray}
\label{eq:Cdef}
{\vec {\cal C}}_{\kappa} =  d {\vec {\cal E}} + \kappa  |g| \mu_0 {\vec {\cal B}}
\end{eqnarray}
with $\kappa = \pm1$.
This sign convention implies that $\kappa =+1$ stands for an ``energetically
stretched'' state.  That is,
for parallel ${\cal E}$ and ${\cal B}$ fields, both fields shift the energy in the same
direction.  Thus the electric and magnetic moments align in the same direction
in the body frame of the molecule, implying in turn that $\lambda + 2\sigma$
is negative.  This means that in general $\kappa$ has the opposite sign to $\omega$.
The geometric construction of the combined fields is illustrated in Fig. 1

\begin{figure}
\begin{center}
\label{Fig1}
\includegraphics[width=5in]{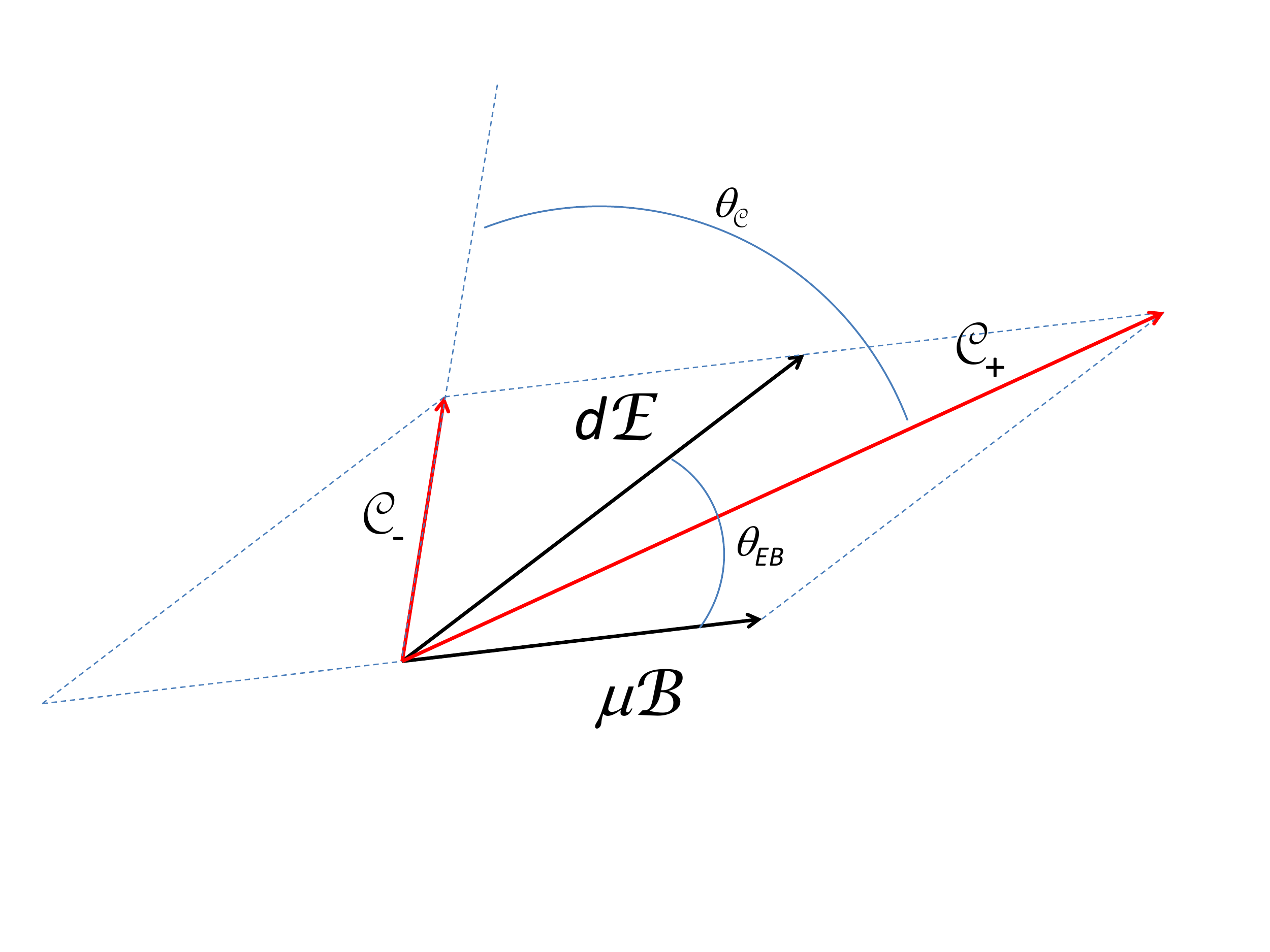}
\end{center}
\caption{Construction of the combined fields in Hund's case-X. Given an electric field
${\vec {\cal E}}$ and a magnetic field ${\vec {\cal B}}$, and the {\it positive} electric
and magnetic moments $d$ and $\mu$, the weighted sums
${\vec {\cal C}}_{\pm} = d{\vec {\cal E}} \pm \mu {\vec {\cal B}}$ describe
appropriate quantization axes for states of a paramagnetic, dipolar case-a radical.   }
\end{figure}

An alternative sign convention would give $\kappa$ the same sign as
$\omega = \lambda + \sigma$ itself. To make this identification in all
cases would be, however, potentially ambiguous: in a $^3\Pi_0$
state, for instance, $\omega=0$ and cannot serve as a signed quantum number (even though
$g$ is nonvanishing).  Likewise,
the signed value of $g = \lambda + 2\sigma$ is not necessarily helpful, as it is
nominally zero for states such as $^2 \Pi_{1/2}$.  The actual $g$-factor of course
can be nonzero, but its sign can be difficult to determine
without detailed consideration of the molecule.  For these reasons, $\kappa$ emerges as a new quantum number,
with obvious ties to $\omega$, that is nevertheless distinct from it.

For a particular internal state identified by  $\lambda$ and $\sigma$ (therefore,  $\kappa$ is determined), ${\vec {\cal C}}_{\kappa}$ serves as a quantization axis.  By analogy with the above, the Hamiltonian becomes
\begin{eqnarray}
\langle \lambda \sigma | \langle \omega j m_{\kappa} |
\left( H_{\rm S} + H_{\rm Z} \right)
|\omega j m_{\kappa} \rangle | \lambda \sigma \rangle =
-{\cal C}_{\kappa} \langle \omega j m_{\kappa} | \cos (\beta) | \omega j m_{\kappa} \rangle,
\end{eqnarray}
where {\it this} $\beta$ is the angle between ${\hat n}$ and ${\vec {\cal C}}_{\kappa}$,
and the subscript on $m_{\kappa}$ emphasizes the angular momentum projection onto the ${\hat {\cal C}}_{\kappa}$ axis.
From here, the problem is mathematically equivalent to the results above.  In particular, the exact energy spectrum of the field Hamiltonian is given by
\begin{eqnarray}
\label{crossed}
H_S + H_Z = - \left( \frac{ \omega }{ \sqrt{ j(j+1)} } \right)
\left( {\cal C}_{\kappa} \frac{ m_{\kappa} }{ \sqrt{j(j+1)} } \right),
\end{eqnarray}
where ${\cal C}_{\kappa}$ is the magnitude of the combined field,
\begin{eqnarray}
\label{eq:combined}
{\cal C}_{\kappa} = \sqrt{ (d{\cal E})^2 + (g \mu_0 {\cal B})^2
+ 2 \kappa d |g| \mu_0 {\cal E} {\cal B} \cos(\theta_{EB}) },
\end{eqnarray}
and $\theta_{EB}$ is the angle between the electric and magnetic fields.
Significantly, (\ref{crossed}) is diagonal in this basis, {\it provided}
that $m_{\kappa}$ refers to quantization along the field axis ${\hat C}_{\kappa}$
{\it for a particular value of} ${\kappa}$.  In this sense $\kappa$ denotes another quantum number of the combined field-molecule system, on which others are contingent.

The crossed-field case can therefore be solved exactly, and quantum numbers
can be assigned to the different energy levels, for electric and magnetic fields
of arbitrary strength and relative orientation.  The way to make this possible is to
accept that the quantum numbers are now conditional, that is, $m_{\kappa}$ cannot be assigned
unambiguously until the internal state $\omega$ (and hence
$\kappa $) are specified.   Because these states identify good quantum numbers in the crossed
field case (suggested by the letter  ``X''), we refer to this basis as the Hund's case-X coupling scheme,
with basis sets denoted $| \omega \kappa j m_{\kappa} \rangle$.

\subsection{Lambda doubling}

We have deliberately focused on the situation were the electric field interaction, $d{\cal E}$, is larger than the $\Lambda$-doublet splitting $\Delta$ in the molecule.  This has ensured that the signed values of $\omega$ are good quantum numbers, rather than the parity quantum number $\epsilon_p$ in the parity states
$\left( |{\bar \omega} \rangle +\epsilon_p |-{\bar \omega} \rangle \right)/\sqrt{2}$.  To complete the picture, we must construct matrix elements of $\Lambda$-doubling in our basis.

Starting with the case of zero magnetic field, the Hamiltonian consists of Stark
and $\Lambda$-doubling terms
\begin{eqnarray}
H = H_{\rm S} + H_{\Lambda}.
\end{eqnarray}
In the signed basis $|\pm {\bar \omega} j m   \rangle$ the Hamiltonian matrix reads
\begin{eqnarray}
H = \left( \begin{array}{cc}
-d {\cal E} \frac{ m {\bar \omega} }{ j(j+1) } & \frac{ \Delta }{ 2 } \\
\frac{ \Delta }{ 2 } & + d {\cal E} \frac{ m {\bar \omega} }{ j(j+1) }
\end{array} \right),
\end{eqnarray}
which gives the familiar eigenvalues
\begin{eqnarray}
\frac{ \omega }{ {\bar \omega} } \sqrt{ \left( d {\cal E} \frac{ m {\bar \omega} }{ j(j+1) } \right)^2
+\left( \frac{ \Delta }{ 2 } \right)^2 },
\end{eqnarray}
where $\Delta$ is the zero-field $\Lambda$-doublet splitting.
More concisely, nonzero matrix elements of the $\Lambda$-doubling Hamiltonian are given by
\begin{eqnarray}
\langle -\lambda -\sigma | \langle -\omega j m^{\prime} | H_{\Lambda}
| \omega j m \rangle | \lambda \sigma  \rangle
= \frac{ \Delta }{ 2 } \delta_{m^{\prime} m},
\end{eqnarray}
provided $m^{\prime}$ and $m$ are referred to the same quantization axis.
This form of the Hamiltonian makes evident that the $\Lambda$-doubling connects
states of $+{\bar \omega}$ to states of $-{\bar \omega}$, that is, in the case-X picture
it mixes the states
$\pm \kappa$ that refer to different axes ${\vec {\cal C}}_{\kappa}$.

To compute matrix elements of $H_{\Lambda}$ in the case-X basis, we therefore have to transform between these two axes.  For concreteness, denote by $m_{\pm}$ quantum numbers referred to the ${\cal C}_{\pm }$ axis, and let
$\Omega_{\cal C} = (0, \theta_{\cal C}, 0)$ be the set of Euler angles
defining the rotation between these axes, with
$\cos (\theta_{\cal C}) = {\hat {\cal C}}_{+} \cdot {\hat {\cal C}}_{-}$.  Without loss of generality, the plane of the two axis defines the laboratory $x$-$z$ plane, whereby the other two Euler angles can be set to zero.  The rotation matrix between the two axes
is then denoted $D$, with matrix elements
$D^j_{m_- m_+}(0, \theta_{{\cal C}},0) = d^j_{m_- m_+}( \theta_{{\cal C}})$
in terms of the Wigner $D$ matrices \cite{BS}.

Writing the Hamiltonian in block-diagonal form, with the blocks denoting $\kappa = +1$
and $\kappa = -1$ states, the transformation reads
\begin{eqnarray}
\left( \begin{array}{cc} 1 & 0 \\ 0 & D^{\dagger} \end{array} \right)
\left( \begin{array}{cc} 0 & H_{\Lambda} \\ H_{\Lambda} & 0 \end{array} \right)
\left( \begin{array}{cc} 1 & 0 \\ 0 & D \end{array} \right)
= \left( \begin{array}{cc} 0 & H_{\Lambda}D \\  D^{\dagger} H_{\Lambda} & 0
\end{array} \right),
\end{eqnarray}
whereby the matrix elements of $\Lambda$-doubling in the case-X basis read
\begin{eqnarray}
\langle -\lambda -\sigma | \langle -\omega \kappa  j m_{\kappa}
| H_{\Lambda} |
\omega -\kappa j m_{-\kappa} | \lambda \sigma \rangle =
\frac{ \Delta }{ 2 } d^j_{m_{\kappa } m_{-\kappa }}( \theta_{{\cal C}}).
\end{eqnarray}
Matrix elements between states with the same value of $\omega$ (hence connecting states from the same set $m_-$ or $m_+$) vanish.

\section{Examples and applications}

Armed with these analytic results, certain aspects of molecules in the combined fields can be elucidated.

\subsection{Magnetic trapping of polar radicals}

Cold paramagnetic radicals are amenable to magnetic trapping in mangetostatic traps,
just as, say, alkali atoms are.  In principle, this would leave the electric field as an independently variable tool to manipulate and study the electric field response of these
dipolar species.  However, for case-a molecules like OH, electric and magnetic
field effects are confounded, as detailed above.  Within the case-X formalism,
we can identify the states involved for a particular field configuration.

\begin{figure}
\begin{center}
\includegraphics[width=5in]{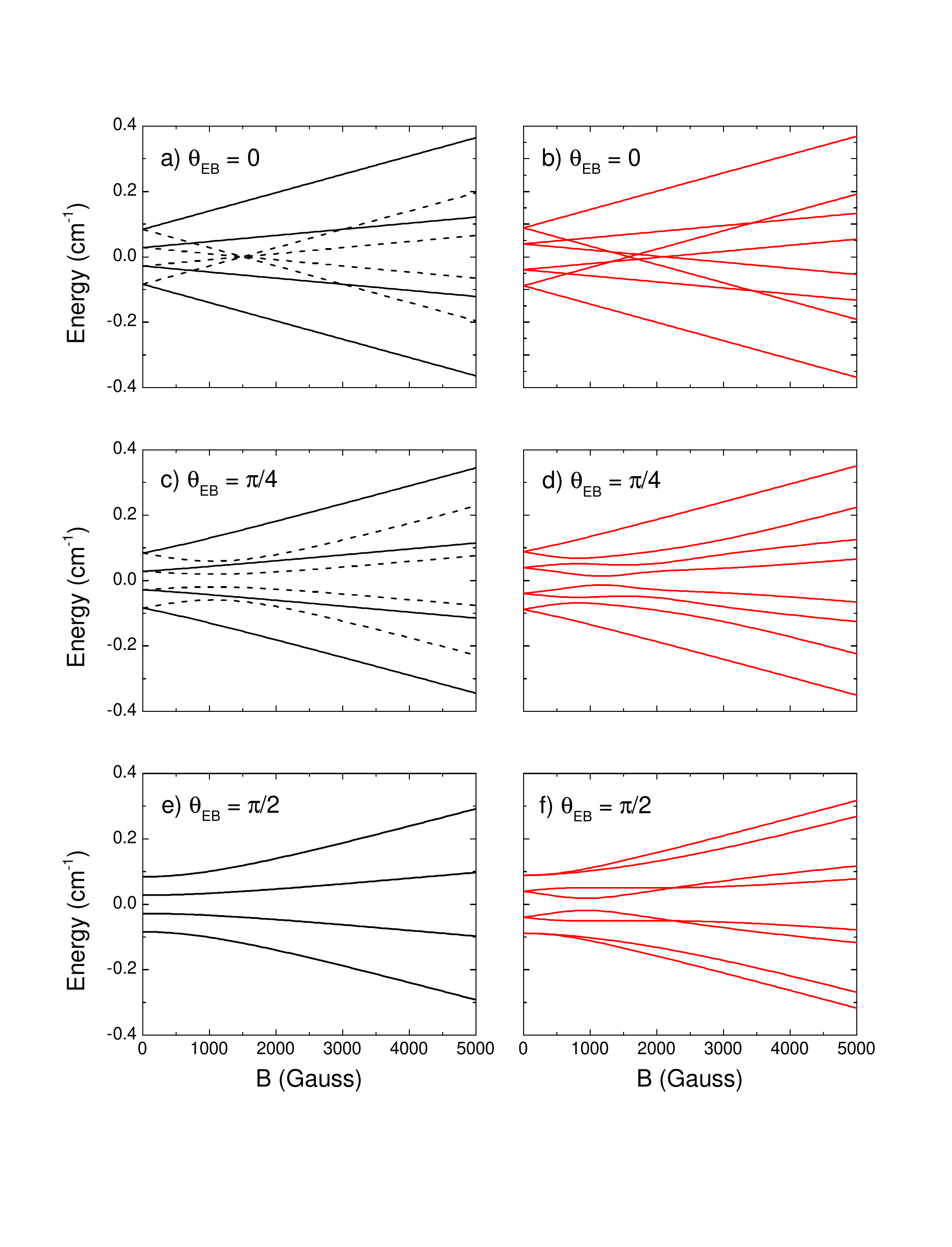}
\end{center}
\caption{Zeeman effect for OH molecules in their $^2 \Pi_{3/2}$, $j=3/2$ ground state,
subject also to an electric field of magnitude ${\cal E}=5$ kV/cm that makes an
angle $\theta_{EB}$ with respect to the magnetic field.  Shown is the approximation
without including $\Lambda$-doubling (left column, black), and including it (right column, red).
In each panel on the left, states with $\kappa = +1$ are drawn using solid black lines,
while those with $\kappa = -1$ are drawn using dashed black lines.
}
\end{figure}

For example, for collinear fields, $\theta_{EB} = 0$, the energies of the states read
\begin{eqnarray}
- \frac{ m \omega }{ j(j+1) }\left( d{\cal E} +
\kappa |g| \mu_0 {\cal B} \right).
\end{eqnarray}
This formula emphasizes the fact that, for $\kappa>0$, the magnetic dipole moment
points parallel to the electric dipole moment.  Thus those states that rise in an electric
field rise further in a magnetic field, and those that decrease in an electric field
decrease further in a magnetic field, an effect which has long been known
\cite{Coester50_PR}.  This is shown in the plot of energies versus
${\cal B}$ in Figure 2a, where ${\cal B}$ is parallel to an electric field
of magnitude ${\cal E}=5$ kV/cm, for the ground $^2 \Pi_{3/2}(j=3/2)$ state of OH.
The solid lines denote the $\kappa = +1$
states.  By contrast, for the $\kappa = -1$ states (dashed lines), the magnetic and
electric fields pull in opposite directions; higher-energy states go lower, and
vice versa, leading to a crossing at $\sim 1500$ Gauss.  To finish off the picture of parallel fields, Fig. 2b
shows the same energy levels but including the effect of $\Lambda$-doubling (computed
in the case-X basis as described above), illustrating
that, while it breaks the degeneracy, it is indeed a perturbation.

Fig. 2c shows the energies of the same OH molecule in the same 5 kV/cm electric
field, but with a magnetic field tilted at an angle $\theta_{EB} = \pi/4$
relative to it.  Again the $\kappa  = +1$ states are denoted by solid lines, while
the $\kappa = -1$ states are denoted by dashed lines.  In this case the
combined field ${\vec {\cal C}}_{+}$  lies somewhere in the acute
angle between the directions of ${\vec {\cal E}}$ and ${\vec {\cal B}}$
(Fig. 1).  Thus, while the fields don't pull the molecular axis in quite
the same direction, they almost do so.  The effect is that the $\kappa = +1$ states
fan out in energy, just as in Fig 2a.

More interesting are the $\kappa = -1$ states (dashed lines) in Figure 2c.  The degeneracy
that was apparent for parallel fields in 2a is now gone, replaced by what
appear to be avoided crossings.  These crossings can be considered as the effect of
the magnetic field breaking rotational symmetry about ${\vec {\cal E}}$ and therefore
mixing states of different $m$ referred to ${\vec {\cal E}}$.  However, in the
case-X picture, each of these states is still characterized by
a unique value of the quantum number $m_-$ for any value of ${\cal B}$.
The quantization axis is of course different for each
${\cal B}$ -- but, it is exactly determined by (\ref{eq:Cdef}).

As before, the effect of including the $\Lambda$-doubling is to perturb
these energies somewhat (Fig. 2d).   Doing so of course introduces couplings between
the case-X states and generates true avoided crossings.  Still, away from the main region of crossings, the case-X quantum numbers $(\kappa, m_{\kappa})$ remain useful
for classifying states.  Approximate energy eigenvalues for this situation,
for spin-1/2 molecules, were extracted from a semiclassical model in
Ref. \cite{Lara08_PRA}.

Finally, consider the case where the electric and magnetic fields are
perpendicular, Fig. 2e.  Now the two combined fields ${\vec {\cal C}}_{\pm}$
point in opposite directions, but lie along the same line.
As a consequence, they define the same quantization axis.  From Eqns. (\ref{crossed},\ref{eq:combined}) it can be seen that, in the absence of
$\Lambda$-doubling, each state $(\kappa = +1, m_{+})$ is exactly
degenerate with the state $(\kappa = -1, m_{-} = -m_{+})$.  Including the
$\Lambda$-doubling therefore mixes degenerate states of opposite parity at all values of ${\cal B}$,
hence has a comparatively large influence on
the spectra even at large fields (Figure 2f).

Incorporating an electric field into the OH magnetic trap also has implications for
Majorana transitions in the trap.  Consider the magnetic field configuration of
a quadrupole trap,
\begin{eqnarray}
{\vec {\cal B}}({\vec r})
= \delta {\cal B} ( x {\hat x} + y {\hat y} - 2z {\hat z}),
\end{eqnarray}
where $\delta {\cal B}$ represents the field gradient.  A magnetic moment ${\vec \mu}$
that adiabatically tracks this field, and is everywhere parallel to it, experiences
a trapping potential
\begin{eqnarray}
U_{\rm trap}({\vec r}) = -{\vec \mu } \cdot {\vec {\cal B}}({\vec r})
= \mu \delta {\cal B} ( x^2 + y^2 +4z^2)^{1/2}.
\end{eqnarray}
The problem, of course, is that the magnetic moment cannot track the field adiabatically
at ${\vec r} = 0$, where the field vanishes.  This nonadiabaticity leads to
the Majorana losses.

In combined fields, however, the situation is different.  Suppose, for example, that the
electric field ${\vec {\cal E}} = ({\cal E}_x,{\cal E}_y,{\cal E}_z)$ is constant in the
vicinity of ${\vec r}=0$.  Then the combined fields
\begin{eqnarray}
{\vec {\cal C}}_{\pm} =
(d{\cal E}_x \pm |g|\mu_0 \delta {\cal B} x){\hat x} +
(d{\cal E}_y \pm |g|\mu_0 \delta {\cal B} y){\hat y} +
(d{\cal E}_x \mp 2|g|\mu_0 \delta {\cal B} z){\hat z}
\end{eqnarray}
do not vanish at ${\vec r}=0$.  Now, assuming that the molecular orientation
${\hat n}$ can adiabatically follow the fields, we have confining potentials
\begin{eqnarray}
U_{\rm trap,\pm}({\vec r}) = \left[
(d{\cal E}_x \pm |g|\mu_0 \delta {\cal B} x)^2 +
(d{\cal E}_y \pm |g|\mu_0 \delta {\cal B} y)^2 +
(d{\cal E}_z \mp 2 |g|\mu_0 \delta {\cal B} z)^2 \right]^{1/2}
\end{eqnarray}
This rounding out of the trap minimum may be expected to reduce the rate of Majorana losses.

\subsection{Electric dipole moments}

In some applications, notably cold collisions, the electric dipole moment
and its orientation play a decisive role.  In a case-X state, the dipole moment
${\vec d} = d {\hat n}$ precesses about the appropriate quantization axis
${\hat {\cal C}}_+$ or ${\hat {\cal C}}_-$ just as it would about the
electric field axis in the absence of a magnetic field.  The semiclassical
direction of the mean dipole $\langle {\vec d} \rangle$ is therefore
unambiguously defined along one of these axes.  In applications, however,
it may also be useful the relate this direction to the direction of the
electric field, to anticipate the role of a magnetic field in re-orienting
$\langle {\vec d} \rangle$ in the lab frame.

To this end, we define the tilt angle,  $\theta_{\rm tilt}$, between
$\langle {\vec d} \rangle$ and ${\vec {\cal E}}$, given by
\begin{eqnarray}
\cos (\theta_{\rm tilt}) &=&
\frac{ (d{\vec {\cal E}}) \cdot {\vec {\cal C}}_{\kappa} }
{ |d{\vec {\cal E}}| |{\vec {\cal C}}_{\kappa}| } \\
&=& \frac{ d{\cal E} + \kappa |g| \mu_0 {\cal B} \cos(\theta_{EB}) }
{ \sqrt{ (d{\cal E})^2 + (g \mu_0 {\cal B})^2
+ \kappa d|g|\mu_0 {\cal E} {\cal B} \cos(\theta_{EB}) } }
\end{eqnarray}
This tilt angle is shown in Fig. 3 versus electric field, for OH in a ${\cal B}=1000$
Gauss magnetic field (and neglecting $\Lambda$-doubling).  The angle between
the fields is arbitrarily set at $\theta_{EB} = \pi/3$.  At large
electric field, $\theta_{\rm tilt}$ goes to zero; a strong electric field
of course polarizes the dipole along itself, regardless of the magnitude and
orientation of the magnetic field.  For smaller electric fields, the
magnetic field makes a significant difference in the dipole's direction.
In the limit of zero electric field, it is rather the magnetic field that sets
the direction of the electric dipole.  In this limit the angle between
$\langle {\vec d} \rangle$ and ${\vec {\cal E}}$ approaches
$\cos (\theta_{\rm tilt}) = \kappa \cos( \theta_{EB})$.   That is,
in this limit $\theta_{\rm tilt} = \theta_{EB}$ when $\kappa = +1$,
and $\theta_{\rm tilt} = \pi - \theta_{EB}$ when $\kappa = -1$.

\begin{figure}
\begin{center}
\includegraphics[width=4in]{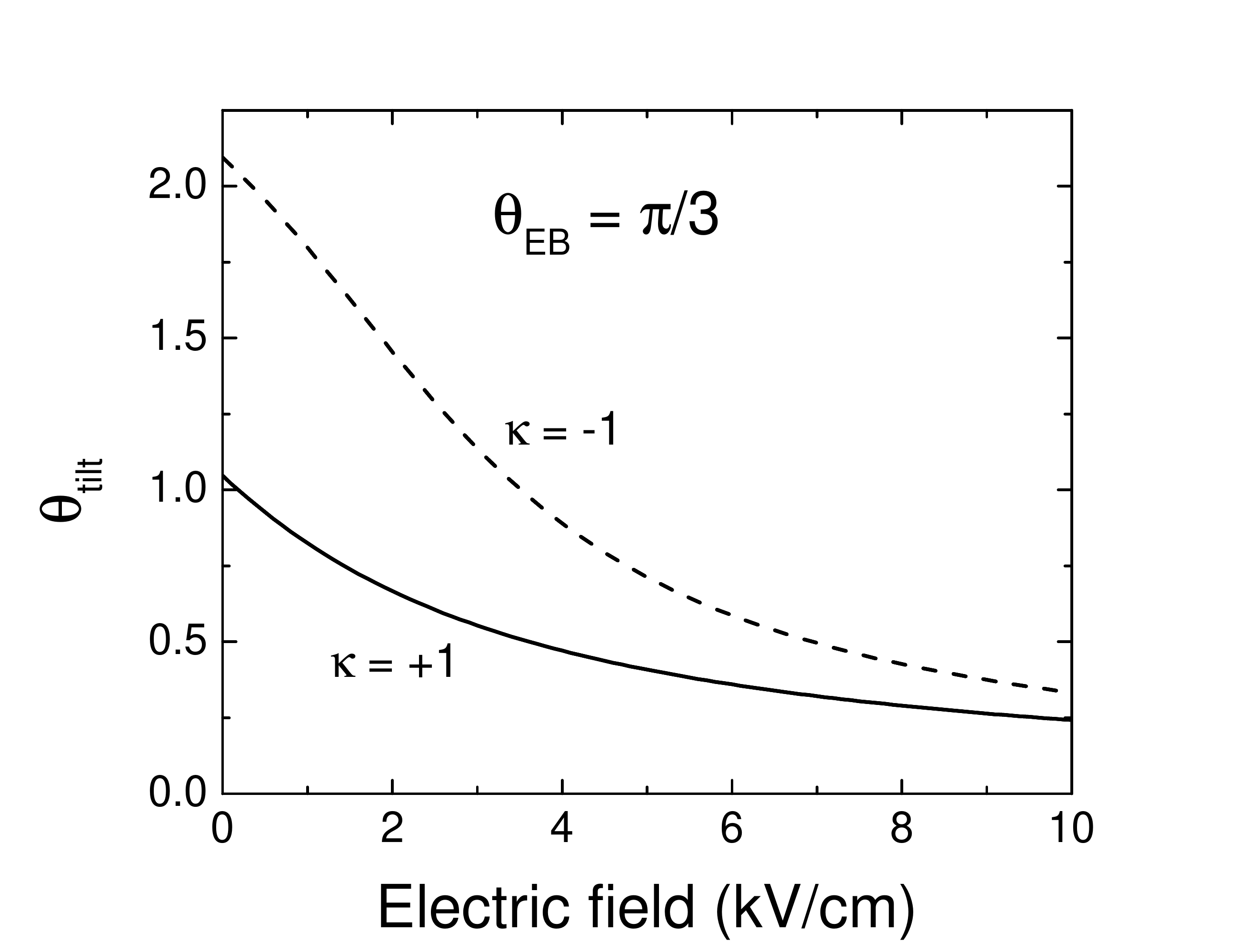}
\end{center}
\caption{The tilt angle $\theta_{\rm tilt}$ between an applied electric field and
the dipole moment of an OH radical, as a function of electric field.  It is assumed
that there is also a magnetic field applied, of strength ${\cal B}= 1000$ Gauss, and making an angle $\theta_{EB} = \pi/3$ with respect to the electric field, and that there is no $\Lambda$-doubling. Solid and dashed
lines refer to $\kappa = \pm 1$ states, respectively.
}
\end{figure}

\subsection{Higher fields and pendular states}

At electric fields sufficiently high that the Stark energy $d {\cal E}$
becomes comparable to, or larger than, the rotational constant $B_e$,
the energy level spectrum qualitatively changes.  In the extreme limit of
$d {\cal E} / B_e \gg 1 $, the molecule is better described as a two-dimensional
harmonic oscillator, which description serves as a starting point for
perturbatively evaluating energies at finite values of $d {\cal E} / B_e \gg 1 $
\cite{Peter57_JCP} (In other words, the oscillator quantum numbers describe the Hund's case
appropriate in the high-filed limit).  These hindered rotor states, dubbed ``pendular states''
\cite{Friedrich91_Nat,Rost92_PRL},
have been explored experimentally for both the electric field
and magnetic field \cite{Slenczka94_PRL} versions, as well as in combined fields that
are either parallel or antiparallel \cite{Friedrich94_CPL}.

Here we merely point out that the case-X classification scheme serves to identify
energy levels even in the event that the fields are non-parallel.  To this end, the
combined field Hamiltonian takes its full, $j$-mixing form
\begin{eqnarray}
H_{\rm S} + H_{\rm Z} &=& -{\cal C}_{\kappa} \langle \omega \kappa j^{\prime}
m_{\kappa} |
\cos(\beta) | \omega \kappa j m_{\kappa} \rangle \\
&=& -{\cal C}_{\kappa}
(-1)^{m_{\kappa} - \omega} \sqrt{ (2j^{\prime}+1)(2j+1) }
\left( \begin{array}{ccc} j^{\prime} & 1 & j \\
                          -\omega & 0 & \omega \end{array} \right)
\left( \begin{array}{ccc} j^{\prime} & 1 & j \\
                          -m_{\kappa} & 0 & m_{\kappa} \end{array} \right). \nonumber
\end{eqnarray}
Significantly, the definition of the combined fields ${\vec {\cal C}}_{\kappa}$,
and the subsequent conservation of $m_{\kappa}$'s along these axes, is
independent of the fact that $j$ is not conserved in a field.  The quantum numbers $\kappa$
remain as good as before.  To the  field interaction, we add the rotational Hamiltonian
\begin{eqnarray}
H_{\rm rot} = ({\vec j}^2 - \omega^2) B_e,
\end{eqnarray}
and diagonalize in a suitable basis of $j$ to determine the energy levels.

\begin{figure}
\begin{center}
\includegraphics[width=4in]{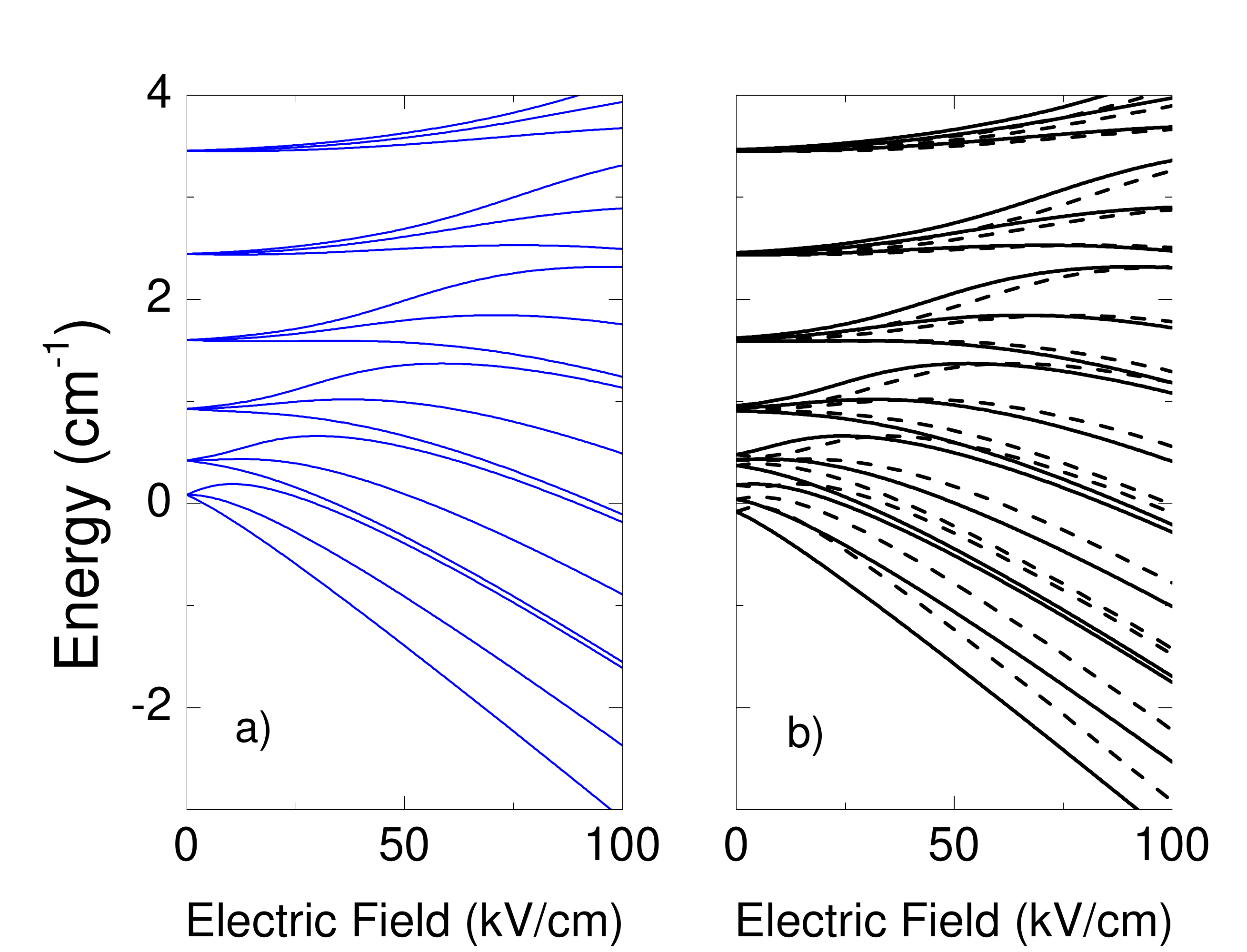}
\end{center}
\caption{Stark effect for ICl molecules in their $A ^3 \Pi_1$ state.  In a) is shown the
energies in the absence of a magnetic field (thin blue lines).  In b), a magnetic field
${\cal B} = 3000$ Gauss is applied, which makes an angle $\theta_{EB}=\pi/4$ with respect to
the electric field.  Solid and dashed lines refer to $\kappa = \pm 1$ states, respectively. }
\end{figure}

This procedure is carried out, with results shown in Fig. 4 for the $A ^3\Pi_1$
state of the ICl molecule \cite{Friedrich94_CPL}.  This figure displays the low-lying
energies, versus electric field, over a range that shows the transformation
between rotor and pendular states, for the low-lying states.  The light, blue line
in a) is the result in zero magnetic field.  In the presence of a magnetic field
${\cal B}= 3000$ Gauss, tilted at an angle $\pi/4$ with respect to the electric field,
degeneracies are broken, leading to independent spectra in 4b) for $\kappa = +1$ (solid black)
and $\kappa = -1$ (dashed black) states.

\section{Conclusion}

In the presence of crossed electric and magnetic fields, neither field alone
serves as a suitable quantization axis for eigenstates of a case-a molecule.
Interestingly, quantization axes can nevertheless be found, and good
quantum numbers $m$ defined for the crossed-field situation.  The cost
of being able to do so is that {\it two} quantization axes must be identified,
which naturally divides the eigenstates into two qualitatively different varieties,
according to whether the electric and magnetic dipole moments are parallel
or antiparallel.  These axes in general also depend upon the electronic
state of the molecule through the quantum numbers $\lambda$ and $\sigma$.

Finally, we remark that the combined fields are not necessary for molecules
described by Hund's case-b.  For these molecules, the electronic spin
is sufficiently decoupled from the molecular axis ${\hat n}$ that the usual
laboratory-frame quantum numbers can be used.  Specifically, the states $|sm_s \rangle$
of spin and $| n m_n \rangle$ of rotation will diagonalize the
Hamiltonian $H_{\rm S} + H_{\rm Z}$, provided that $m_s$ is quantized along
the magnetic field axis, and $m_n$ along the electric field axis.

This material is based upon work supported by the Air
Force Office of Scientific Research under the Multidisciplinary
University Research Initiative Grant No. FA9550-09-1-0588.

\end{document}